\documentclass{appolb}
\usepackage{amsmath,amssymb,amsthm,bm,bbm,color,hyperref}
\usepackage{graphicx,subfigure,rotating}

\begin{document}

\title{Center Vortex vs. Abelian models of the QCD vacuum} 

\author{Roman H\"ollwieser$^a$$^c$\footnote{Speaker at the 4th
winter workshop on Non-Perturbative Quantum Field Theory, February 2-5, 2015,
Sophia Antipolis, Nice, France. Funded by an Erwin Schr\"odinger Fellowship of 
the Austrian Science Fund under Contract No. J3425-N27.}, and
Jeff Greensite$^b$\footnote{Supported by
the U.S.\ Department of Energy under Grant No.\ DE-FG03-92ER40711.}
\address{$^a$Institute of Atomic and Subatomic Physics, Nuclear Physics Dept.\\
Vienna University of Technology, Operngasse 9, 1040 Vienna, Austria},
\address{$^c$Department of Physics, New Mexico State University, \\
Las Cruces, NM, 88003-0001, USA}
\address{$^b$Physics and Astronomy Department, San Francisco State
University,\\ San Francisco, CA~94132, USA}}

\maketitle

\begin{abstract}
We present evidence that the center vortex model of confinement is more
consistent with lattice results than other currently available models. In
particular we show that Abelian field distributions predicted by monopole
plasma, caloron gas or dual superconductor models cannot reproduce the area-law
falloff of double winding Wilson loops in full $SU(2)$ and center vortex only
gauge fields.
\end{abstract}



\section{\label{intro}Introduction}

Quantum chromodynamics (QCD) at low energies is dominated by the
non-perturbative phenomena of quark confinement and spontaneous chiral symmetry
breaking ($\chi$SB). Center vortices are promising candidates 
for explaining confinement. They form closed magnetic flux tubes, whose flux is 
quantized, taking only values in the center of the gauge group. These properties 
are the key ingredients in the vortex model of confinement, which is theoretically 
appealing and was also confirmed by a multitude of numerical calculations, in 
lattice Yang-Mills theory, see~\cite{Greensite:2003bk} and references therein
and within infrared effective models of random center vortex lines in continuous 3D
space-time~\cite{Hollwieser:2014lxa} and world-surfaces in discrete
4D lattices~\cite{Engelhardt:1999wr,Engelhardt:2003wm}. 

Lattice QCD simulations indicate further that vortices are responsible for the
spontaneous breaking of chiral symmetry ($\chi$SB), dynamical mass generation
and the axial $U_A(1)$ anomaly~\cite{deForcrand:1999ms,Leinweber:2006zq,Bornyakov:2007fz,Jordan:2007ff,Hollwieser:2008tq,Bowman:2010zr,Hollwieser:2010mj,Hollwieser:2011uj,Hollwieser:2012kb,Schweigler:2012ae,Hollwieser:2013xja,Trewartha:2015nna}, and thus successfully explain the non-perturbative phenomena which characterize the infrared sector of strong interaction physics.

In these proceedings we present recent results on so called ``double-winding'' Wilson loops, a gauge-invariant observable suitable to test center vortex and Abelian models of confinement by comparison with full SU(2) gauge theory. In view of the ongoing interest in monopole/caloron confinement mechanisms
\cite{Muller-Preussker:2015daa,Shuryak:2014gja,Bruckmann:2011yd,Cea:2014hma,Liu:2015ufa}, it is reasonable to examine those conjectured mechanisms critically. 

\section{\label{models} Abelian fields and Abelian models}

Magnetic monopole confinement mechanisms, in either the monopole plasma
\cite{Polyakov:1975rs,Polyakov:1976fu} or (closely related) dual superconductor
incarnations \cite{Mandelstam:1974pi,'tHooft:1976ff}, provide a durable image of
the mechanism underlying quark confinement in non-Abelian gauge theories.  The
more recent notion that long-range field fluctuations in QCD are dominated by
caloron gas ensembles \cite{Gerhold:2006sk,Diakonov:2007nv}, fits nicely into
the framework of the earlier monopole plasma conjectures. The mechanisms we are discussing have this point in common: there is some choice of gauge in which the  large scale quantum fluctuations responsible for disordering Wilson loops are essentially Abelian, and are found primarily in the gauge fields associated with the Cartan subalgebra of the gauge group. For the SU(2) gauge group, which is sufficient for our purposes, let this Abelian field be the $A_\mu^3$ color component. 
The question we are concerned with is: what do typical configurations drawn from the Abelian field distribution look like?  Do they resemble what is predicted by monopole plasma, caloron gas, and dual superconductor models?  
To be clear, we do not challenge the notion that, in some gauge, most of the confining fluctuations are Abelian in character. The purpose is to subject a qualitative feature of those predicted distributions to a numerical test.
    
\section{\label{dwind}Double-winding Wilson loops}
Let $C_1$ and $C_2$ be two coplanar loops, with $C_1$ lying entirely in the minimal area of $C_2$, which share a point $\vec{x}$ in common.  Consider a Wilson loop in SU(2) gauge theory which winds once around $C_1$ and once, winding with the same orientation, around $C_2$. 
It will also be useful to consider Wilson loop contours in which $C_1$ lies mainly in a plane displaced in a transverse direction from the plane of $C_2$ by a distance $\delta z$ comparable to a correlation length in the gauge theory. We will refer to both of these cases as ``double-winding'' Wilson loops. In both cases we imagine that the extension of loops $C_1,C_2$ is much larger than a correlation length, so in the latter example the displacement of loop $C_1$ from the plane of $C_2$ is small compared to the size of the loops.  Let $A_1,A_2$ be the minimal areas of loops $C_1,C_2$ respectively.  What predictions can be made about the expectation value $W(C)$ of a double-winding Wilson loop, as a function of areas $A_i$?

In the Abelian models summarized in the previous section, the answer for the
displaced loops simply $W(C) = \exp[-\sigma(A_1 + A_2) - \mu P]$, where $P$ is a
perimeter term, equal to the sum of the lengths of $C_1$ and $C_2$. Assuming
that the large scale fluctuations are Abelian in character, we can make the
``Abelian dominance'' approximation
\begin{eqnarray}
W(C) &=& \frac{1}{2} \left\langle \text{Tr} P\exp\left[i\oint_C dx^\mu A^a_\mu
{\sigma^a \over 2} \right] \right\rangle \approx \left\langle \exp\left[ i \frac{1}{2}\oint_C dx^\mu A^3_\mu \right]  \right\rangle \nonumber \\ &=&  \left\langle \exp\left[ i \frac{1}{2}\oint_{C_1} dx^\mu A^3_\mu \right]  \exp\left[ i \frac{1}{2}\oint_{C_2} dx^\mu A^3_\mu \right] \right\rangle \ .
\end{eqnarray}
If loops $C_1$ and $C_2$ are sufficiently far apart, then the expectation value
of the product is approximately the product of the expectation values, {\it i.e.},
\begin{eqnarray}
W(C) &\approx& 2 \left\langle \exp\left[ i \frac{1}{2}\oint_{C_1} dx^\mu A^3_\mu
\right] \right\rangle \left\langle  \exp\left[ i \frac{1}{2}\oint_{C_2} dx^\mu
A^3_\mu \right] \right\rangle
\nonumber \\
&\approx&  \exp[-\sigma(A_1+A_2)] \ ,
\end{eqnarray}
which we refer to as a ``sum-of-areas falloff''. Now, as $\delta z \rightarrow
0$, we would argue that this limit does not really change the sum-of-areas behavior.
Analytical arguments can be found in~\cite{Greensite:2014gra}. The question is
whether this sum-of-areas behavior corresponds to the actual behavior of
double-winding Wilson loops.

In the center vortex picture of confinement, and also in strong coupling lattice
gauge theory, the behavior of the double-winding loops, whether coplanar or
slightly shifted, is $W(C) = \alpha \exp[-\sigma |A_2 - A_1|]$. The same
difference-of-areas law is obtained in $SU(3)$ pure gauge theory, in the vortex
picture and from strong-coupling expansions, for a Wilson loop which winds twice
around loop $C_1$ and once around the coplanar loop $C_2$.
For simplicity, however, we will restrict
our discussion to $SU(2)$, where $W(C)$ picks up a center element $(-1)$ each time
any of the loops $C_1$ or $C_2$ is pierced by a vortex. So the vortex crossing
can only produce an effect if it pierces the minimal area of $C_2$ but not the
minimal area of $C_1$, resulting in a ``difference-of-areas'' falloff. 
A slight shift of loop $C_1$ by $\delta z$ in the transverse direction does
not make any difference to the argument, providing  the scales of $A_1$ and
$A_2$ are so large compared to $\delta z$ that a vortex piercing the smaller area
$A_1$ is guaranteed to also pierce the larger area $A_2$.

\section{\label{MC}Sum or Difference of Areas?}

\begin{figure}[h]
a)\includegraphics[width=0.40\linewidth]{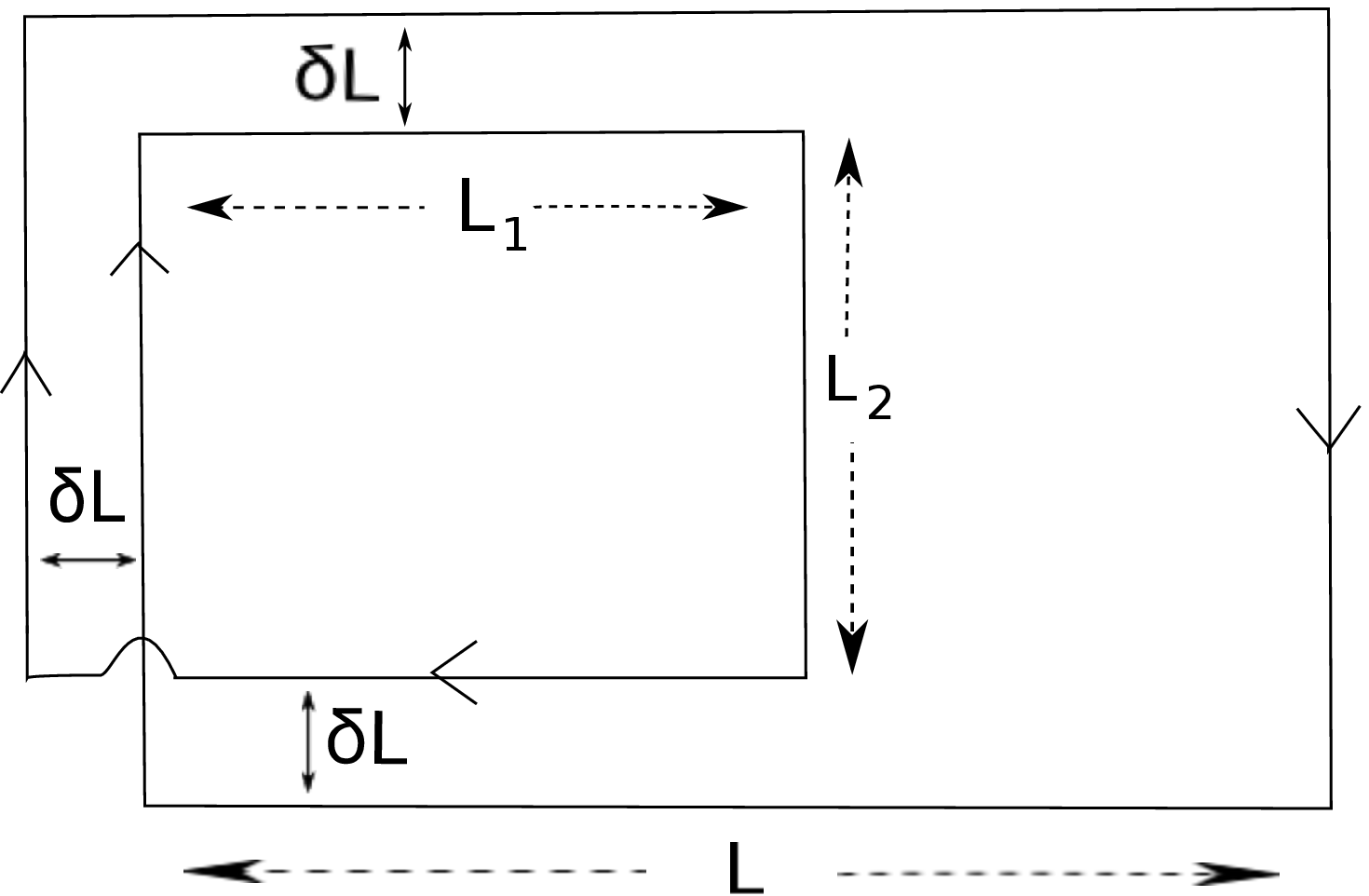}
b)\includegraphics[width=0.46\linewidth]{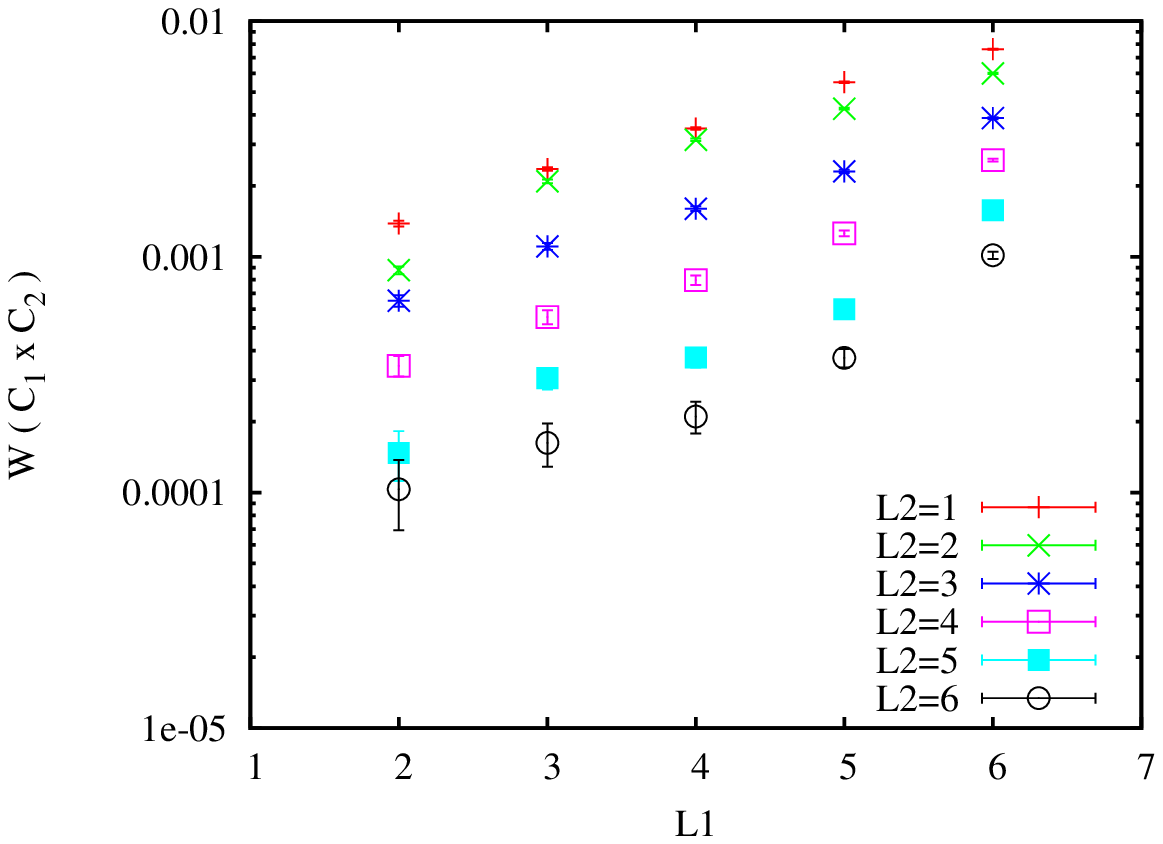}
c)\includegraphics[width=0.46\linewidth]{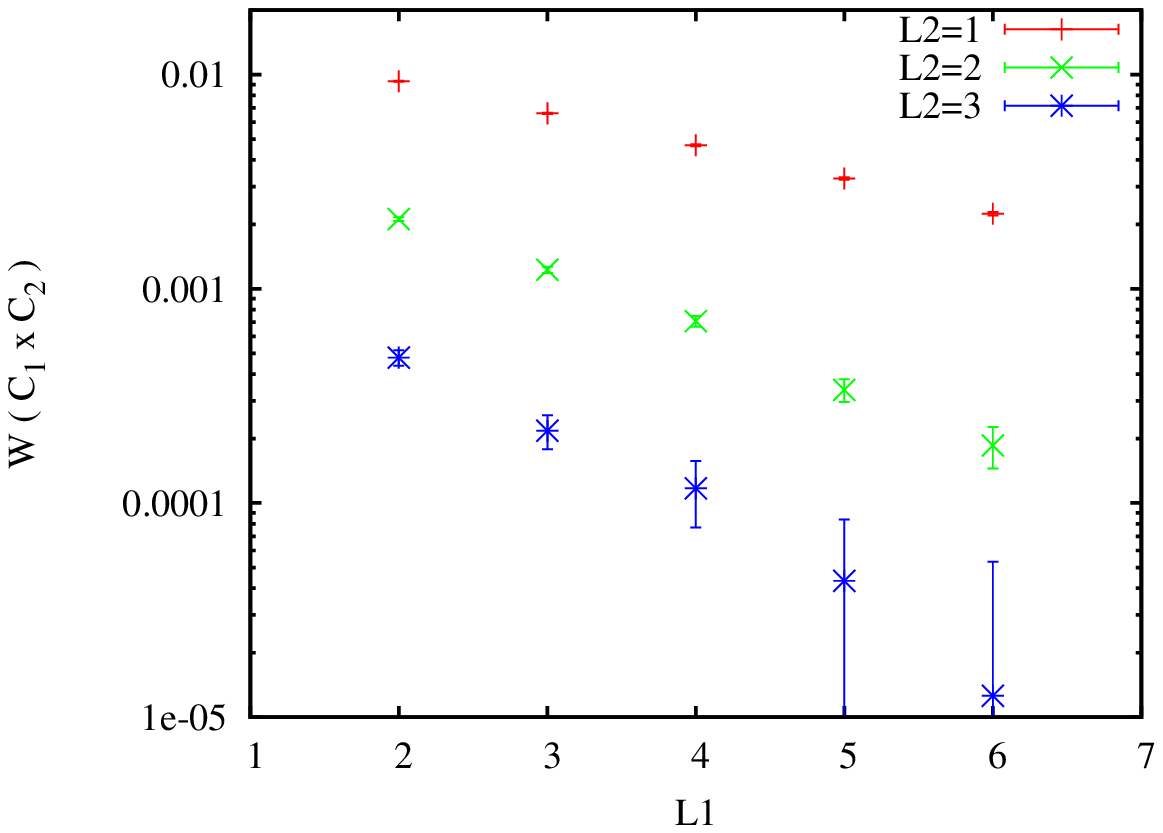}$\;$
d)\includegraphics[width=0.46\linewidth]{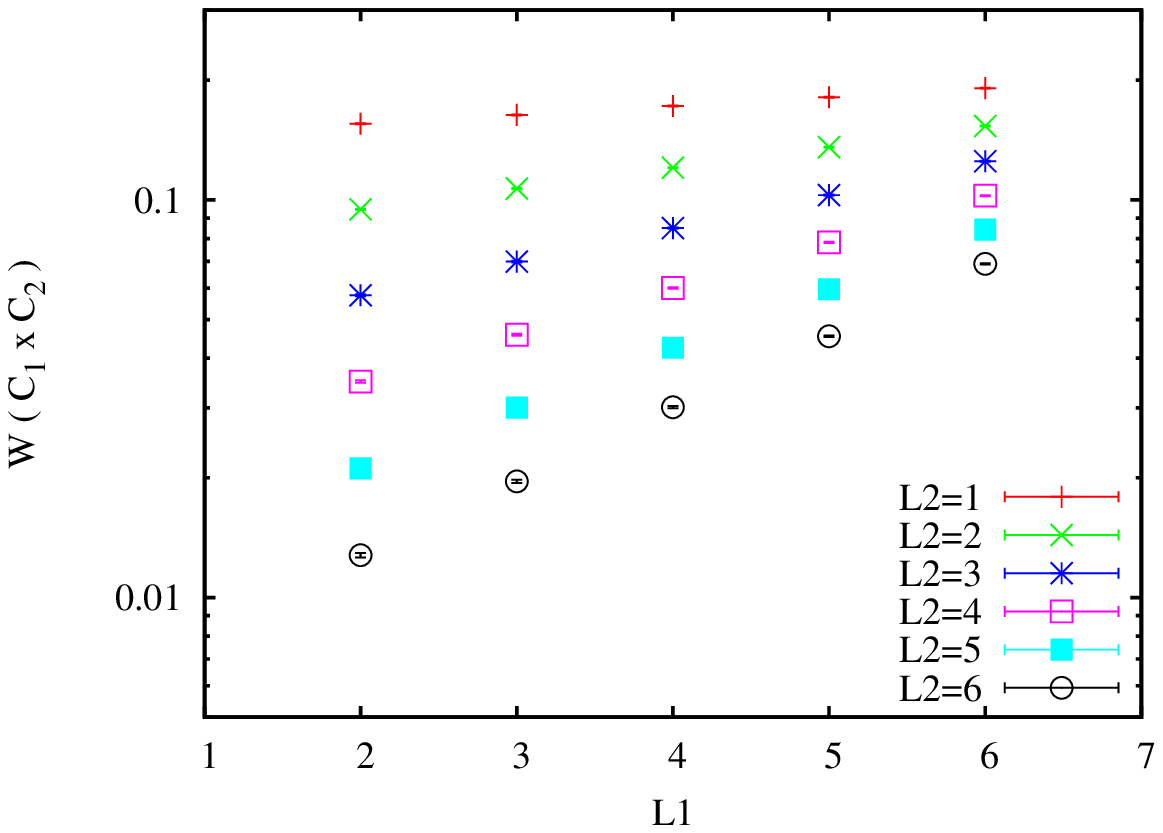}
\caption{a) The coplanar double-winding loop $16^4$ lattices with $L=7$ and $L_2$
fixed and $\delta L=1$. Both unprojected SU(2) loops on smeared links (b), and
center-projected loops in maximal center gauge (d) show a difference-of-areas, 
MAG projected loops (c) show sum-of-areas behavior.}
\label{fig:res}
\end{figure}

\begin{figure}[htb]
 a)\includegraphics[width=.46\linewidth]{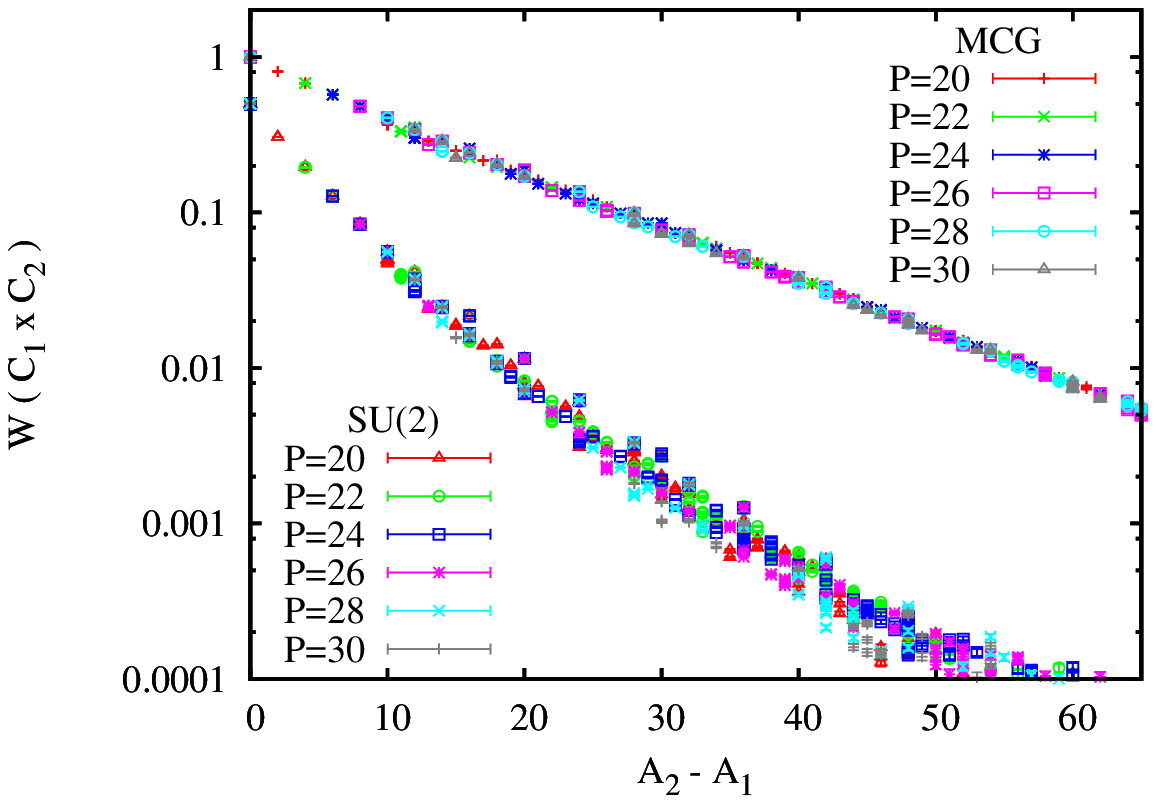}$\;$
 b)\includegraphics[width=.46\linewidth]{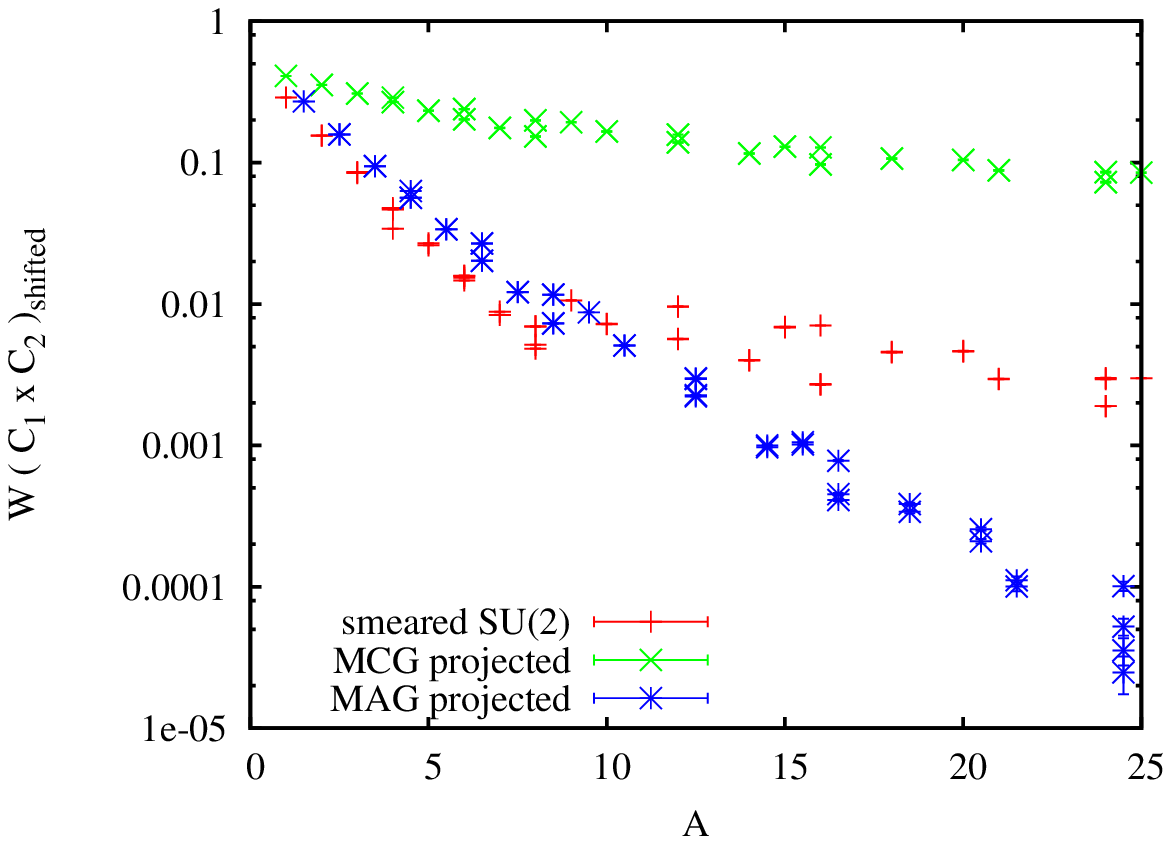}
 \caption{a) Smeared $SU(2)$ and center projected (MCG) results for fixed perimeter $P$ vs. difference-of-areas. b) Double-winding loops $C_1=C_2$ shifted in a  transverse direction by one lattice spacing. The two loops are of equal area $A$, so the difference in area is zero. $W(C)$ for the unprojected SU(2) loops levels off at $A \approx 8$.}
\label{Wshift}
\end{figure}

We consider the contour shown in Fig.\ \ref{fig:res}a, where $\delta L=1, L=7$,
$L_2$ and therefore the size of $C_2$ are fixed and we vary $L_1$, {\it i.e.},
the size of the inner loop $C_1$. If we increase $L_1$ the sum-of-areas (and the
perimeter) increases, however difference-of-areas decreases. Thus if
$W(C_1\times C_2)$ increases with $L_1$, the dominant behavior must be
difference-of-areas, and sum-of-areas is completely ruled out. This is clearly
the case for center-projected loops in maximal center gauge
(Fig.~\ref{fig:res}b) and, more importantly, for gauge-invariant loops as shown
in Fig.~\ref{fig:res}d. The difference-of-areas law is not evident for
Abelian-projected loops in maximal Abelian gauge (Fig.~\ref{fig:res}c). In
Fig.~\ref{Wshift}a we plot results for fixed perimeter $P$, vs.\ the difference
in area $A_2-A_1$ of the contour shown in Fig.~\ref{fig:res}a with $\delta L=0$.
Note that the points seem to cluster around universal lines, regardless of
perimeter. The results clearly show that difference-of-areas is the dominant
effect, and the sum-of-areas behavior is definitely ruled out.

We also look at shifted double-winding loops with $C_1=C_2=C$, so that the
difference in areas is zero. For a transverse shift $\delta z=0$ the situation
is trivial. We can make use of an SU(2) group identity Tr$[U(C) U(C)] = -1 +
\text{Tr}_A U(C)$, where the trace on the right-hand side is in the adjoint
representation.  Since, apart from very small loops, $\langle \text{Tr}_A U(C)
\rangle \ll 1$, we have, almost independent of loop size, $W(C) \approx
-\frac{1}{2}$, which is obviously consistent with difference-in-area behavior,
just like for center-projected loops, where the result is $W(C)=1$ exactly. For
loops $C_1=C_2$ shifted by $\delta z=1a$, where $a$ is the lattice spacing,
there is still almost no effect for center projected loops, and for smeared
$SU(2)$ loops $W(C)$ levels off for large areas $A$, see Fig.~\ref{Wshift}b.
Abelian-projected loops seem to follow a sum-of-areas falloff in this range,
although they may level out eventually.

\vspace{-5mm}  
\section{\label{conclude}Conclusions}

We draw the obvious conclusion that if confinement can be attributed, in some
gauge, to the quantum fluctuations of gauge fields in the Cartan subalgebra of
the gauge group, then the spatial distribution of the corresponding Abelian
field strength cannot follow any of the models discussed in section \ref{models}. 
The difference-of-areas law could be obtained in these models once the neglected
double charged matter, namely W-bosons (or any other double-charged objects in
the model), are properly taken into account. But then the typical distribution
of Abelian fields in the vacuum must be be arranged so as to be consistent with
the difference-of-areas behavior. For example, in a monopole picture, the field
distribution at a fixed time would very likely resemble a chain of monopoles and
anti-monopoles rather than a monopole Coulomb gas, with the magnetic flux
collimated, from monopole to anti-monopole, along the line of the chain. In
other words, rather than being a monopole plasma, this is a vacuum consisting of
center vortices, and the difference-in-area law follows. This is exactly what
happens in compact QED with a double-charged Higgs and numerical evidence for
this picture was also provided in the context of Abelian projection in maximal
Abelian gauge~\cite{Ambjorn:1999ym,Gubarev:2002ek}. In general vortices have a
non-trivial color structure, which in Abelian projection leads to monopole lines
on vortex surfaces and is a key ingredient for chiral symmetry breaking of
center vortices~\cite{Schweigler:2012ae,Hollwieser:2013xja}. 

\bibliographystyle{utphys}
\bibliography{../literatur}

\end{document}